\documentstyle[epsfig,sprocl]{article}
\input{epsf.tex}
\font \fwork = cmssqi8

\newcommand{\nk}{{\bf k}}
\newcommand{\np}{{\bf p}}
\newcommand{\nq}{{\bf q}}
\newcommand{\nJ}{{\bf J}}
\newcommand{\nP}{{\bf P}}
\newcommand{\hp}{{\bf \hat{p}}}

\begin{document}
\hspace*{-18pt}{\epsfxsize=80pt  \epsfbox{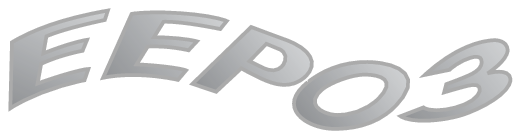}} 
\vspace*{-16pt}
\rightline{\fwork LPSC Grenoble, France, October 14-17, 2003}
\vspace*{15pt}
%
%
\title{\Large\bf Meson Exchange Currents in 
$(\vec{e},e'\vec{p})$ observables
}

\author{\underline{F.K. Tabatabaei$^1$}, J.E. Amaro$^1$, 
J.A. Caballero$^2$, M.B. Barbaro$^3$}

\address{
    {\it $^1$Departamento de F\'{\i}sica Moderna, 
          Universidad de Granada,
          Granada 18071, Spain  \\   
    $^2$Departamento de F\'\i sica At\'omica, Molecular y Nuclear 
          Universidad de Sevilla, Apdo. 1065, 
          Sevilla 41080, Spain \\
    $^3$Dipartimento di Fisica Teorica, Universit\`a di Torino and
INFN, Sezione di Torino 
Via P. Giuria 1, 10125 Torino, ITALY 
}
}    
\maketitle


\section{ Introduction } 

Our main aim in this work is to study the effects of meson-exchange
currents (MEC) over asymmetries and recoil polarization observables in
$A(e,e'p)B$ reactions.  The present DWIA+MEC model\cite{Ama03b,Kaz03}
takes care of relativistic degrees of freedom by using
semi-relativistic (SR) operators for the one-body current as well as
for the two-body MEC.  These SR currents are obtained by a direct
Pauli reduction of the corresponding relativistic operators by an
expansion in terms of missing momentum over the nucleon mass,
$O(p/m_N)$, while the current dependence on the transferred energy and
momentum is treated exactly\cite{Ama02c}.  Relativistic kinematics
for the ejected nucleon is assumed throughout this work. The
final-state interactions (FSI) are incorporated through a
phenomenological optical potential which, for high momentum transfer,
is taken as the Schr\"odinger-equivalent form of a S-V Dirac optical
potential.  The use of the SR model becomes, as a starting point, a
convenient way of implementing relativistic effects in existing non
relativistic descriptions of the reaction mechanism in order to
explore the high momentum region.

\section{ DWIA Model of $(\vec{e},e'\vec{p})$}
We consider the process in which an incident electron with
four-momentum $K^\mu_e=(\epsilon_e,\nk_e)$ and helicity $h$ interacts
with a nucleus $A$, scatters through an angle $\theta_e$ to
four-momentum $K'{}^\mu_e=(\epsilon'_e,\nk'_e)$ and is detected in
coincidence with a nucleon with momentum $\np'=(p',\theta',\phi')$ and
energy $E'$. The four-momentum transferred to the nucleus is
$Q^\mu=K^\mu_e-K'{}^{\mu}_e=(\omega,\nq)$.  We work in the Lab system
with the $z$ axis in the $\nq$ direction and the $x$-axis in the
direction defined by the $\nk_e$-component perpendicular to $\nq$.
The polarization of the final nucleon is measured along an arbitrary
direction defined by the unitary vector $\vec{s}$.  Results will be
shown as  functions of the missing momentum
$\np=\np'-\nq=(p,\theta,\phi)$ for a fixed value of the excitation
energy of the (discrete) final state of the daughter nucleus.
Assuming plane waves for the electrons and neglecting the nuclear
recoil, the cross section can be written in the extreme relativistic
limit
\begin{equation}
\frac{d\sigma}{d\epsilon'_e d\Omega'_e d\hp'}
= \Sigma + h \Delta \, ,
\label{cross-section}
\end{equation}
where a separation has been made into terms involving polarized and
unpolarized incident electrons.  Using the general properties of the
leptonic tensor it can be shown that both terms, $\Sigma$ and
$\Delta$, have the following decompositions:
\begin{eqnarray}
\Sigma 
&=& K \sigma_M \left( v_LR^L+v_T
    R^T+v_{TL}R^{TL}+v_{TT}R^{TT}\right),
\label{sigma}\\
\Delta 
&=&
K \sigma_M \left( v_{TL'}R^{TL'}+v_{T'}R^{T'}\right) \, ,
\label{delta}
\end{eqnarray}
where $\sigma_M$ is the Mott cross section, $K\equiv
m_Np'/(2\pi\hbar)^3$, and the $v_\alpha$-coefficients are the usual
electron kinematical factors.

The hadronic dynamics of the process is contained in the exclusive
response functions $R^K$, in which $K= L,T,TL,TT,T',TL'$.  Isolating
the explicit dependences on the azimuthal angle of the ejected nucleon
$\phi'=\phi$, the hadronic responses can be expressed in the form
\begin{eqnarray}
R^L &=& W^L, \hspace{1cm} 
R^{TL} = \cos\phi\ W^{TL} + \sin\phi\ \widetilde{W}^{TL}, 
\label{R1}\\ 
R^T &=& W^T , \hspace{9mm} 
R^{TT}= \cos2\phi\ W^{TT} + \sin2\phi\ \widetilde{W}^{TT}, 
\label{R2}\\
R^{T'}&=&\widetilde{W}^{T'}, \hspace{7mm} 
R^{TL'}= \cos\phi\ \widetilde{W}^{TL'}+\sin\phi\ W^{TL'} \label{R},
\label{R3}
\end{eqnarray}
where the 9 W-response functions 
$W^K$ and $\widetilde{W}^K$ are totally specified
by four kinematical variables $\{E,\omega,q,\theta'\}$, and the
polarization direction $\{\theta_s,\Delta\phi=\phi-\phi_s\}$.  In the
case of $(\vec{e},e'\vec{N})$ processes, the hadronic response
functions are usually given by referring the recoil nucleon
polarization vector $\vec{s}=(s_l,s_n,s_t)$ 
to the barycentric system defined by the
axes $\vec{l}=\np'/p'$, $\vec{n}=\nq\times\vec{l}/q$, and
$\vec{t}=\vec{n}\times\vec{l}$.  
It can be shown\cite{Pic87} that
the W-response functions  (\ref{R1}--\ref{R3}) can be written as a sum of
unpolarized and spin-vector dependent terms $W^K= \frac12 W_{unpol}^K
+ W_n^K s_n,
\hspace{3mm} K=L,T,TL,TT,TL'$ and $\widetilde{W}^{K}= W_l^{K}s_l +
W_t^{K}s_t, \hspace{3mm} K=TL,TT,T',TL'$. 
Therefore a total of eighteen reduced response functions 
(i.e., independent of the polarization angles) enter
in the analysis of these reactions. 

Alternatively,  the cross
section~(\ref{cross-section}--\ref{delta}) can be written in terms of
the unpolarized one and of the induced and transferred 
polarizations and electron analyzing power 
\begin{equation}
\frac{d\sigma}{d\epsilon'_e d\Omega'_e d\hp'} =
\frac12\Sigma_{unpol}\left[1+\nP\cdot\vec{s}+h( A + \nP'\cdot\vec{s})\right].
\end{equation}

In our model the response functions are obtained by a multipole
expansion following a general procedure developed for 
$(e,e'p)$ reactions from polarized nuclei\cite{Ama98b}. 
We refer to our recent works\cite{Ama03b,Kaz03}  for  more details.

\section{Electromagnetic Operators} 

In the present calculation we consider a semi-relativistic (SR) model for
describing the electromagnetic one-body (OB) and two-body MEC current
operators. The dependence on the transfer and final momenta, which can
be large, is treated exactly.  The SR-OB current, given by
\begin{eqnarray}
J^0(\np',\np)&=&
\rho_c+ i \rho_{so}(\cos\phi\ \sigma_x-\sin\phi\ \sigma_y)\chi,
\\
J^x(\np',\np)&=&
iJ_m\sigma_y + J_c\ \chi\cos\phi,
\\
J^y(\np',\np)
&=&
-iJ_m\sigma_x + J_c\ \chi\sin\phi,
\end{eqnarray}
where $\chi=(p/m_N)\sin\theta$, 
includes charge and spin-orbit parts, in the case of the time-component, and
magnetization and convection parts, in the case of the transverse 
component
\begin{eqnarray}
\rho_c &=& \frac{\kappa}{\sqrt{\tau}}G_E 
\kern 1cm  \rho_{so} = \kappa\frac{2G_M-G_E}{2\sqrt{1+\tau}}
\label{rho-factors}
\\
J_m &=& \sqrt{\tau}G_M  
\kern 1cm
J_c = \frac{\sqrt{\tau}}{\kappa}G_E .
\label{J-factors}
\end{eqnarray}
Here $G_E$ and $G_M$ are the electric and magnetic nucleon form
factors.  Note that this current differs from traditional non
relativistic expansions by relativistic correction factors dependent
on $\kappa=q/2m_N$ and $\tau=|Q^2|/4m_N^2$.

%
\begin{figure}[htb]
\centerline{\epsfig{file=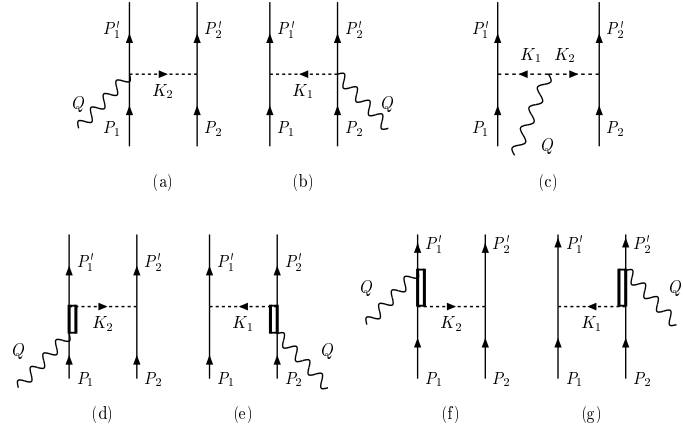,width=9.0cm,height=6.0cm}}
\caption{Feynman diagrams contributing to the two-body current with
one pion-exchange. Contact (C) (a,b), pionic (P) (c), and isobar
($\Delta$) (d)--(g).}
\label{diagram}
\end{figure}
The two-body MEC operators of pionic (P), seagull or contact (S) and
$\Delta$-isobar kinds, displayed in the Feynman diagrams of
Fig.\ref{diagram}, have also been obtained by making use of a SR
approach leading to simple prescriptions that include relativistic
corrections through multiplicative factors:
\begin{equation}
\nJ^{MEC}_{SR} =
\frac{1}{\sqrt{1+\tau}}\nJ^{MEC}_{NR}
\end{equation}
 where $\nJ^{MEC}_{NR}$ is the
traditional non-relativistic MEC operator.

\section{Results.}
\begin{figure}[htb]
\centerline{\epsfig{file=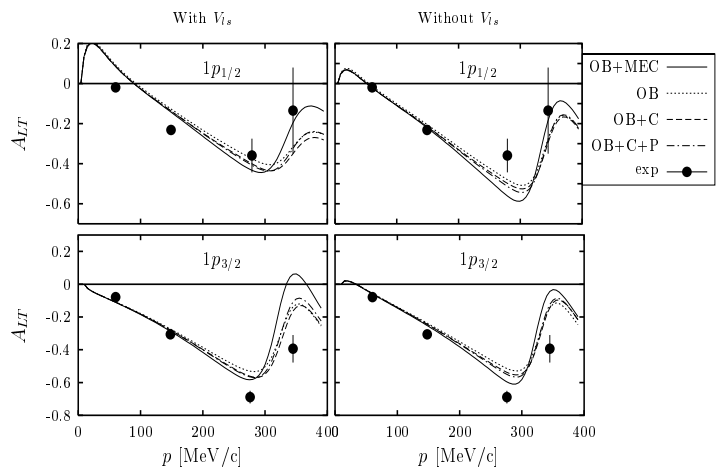,width=9.0cm,height=6.0cm}}
\caption{$A_{TL}$-asymmetry for proton knock-out from the $p$
 shells in $^{16}$O,
for $q=995$ MeV/c and $\omega=439$ MeV,
including the contribution of  MEC.
Experimental data are from~\protect\cite{Gao00}.  Left and right
panels have been obtained including or not the spin orbit part in
the optical potential.}
\label{atl}
\end{figure}

As an example we show results for a selection of observables in proton
knock-out from the $p_{1/2}$ and $p_{3/2}$ shells of $^{16}$O.  In
Fig.\ref{atl} we show results for the $A_{TL}$ asymmetry for $q=995$
MeV/c and $\omega=439$ MeV.  $A_{TL}$ is obtained from the difference
of unpolarized cross sections measured at $\phi'=0$ and
$\phi'=180^{\rm o}$ divided by the sum, hence this observable is
particularly interesting because it does not depend on the
spectroscopic factors.  The effect of MEC is very small for low
missing momentum values and it starts to be important for $p\geq 300$
MeV/c, a region where other relativistic effects are also playing a
role \cite{Udi99}.  We show two sets of calculations: including the
spin-orbit part of the optical potential $V_{ls}$ (left panels) and
without it (right panels). Note the discrepancy between the
calculation and the data for low $p$ in the case of $p_{1/2}$ with the
full potential.  Our results show that the MEC effects for high
missing momenta strongly depend on the FSI.

\begin{figure}[htb]
\centerline{\epsfig{file=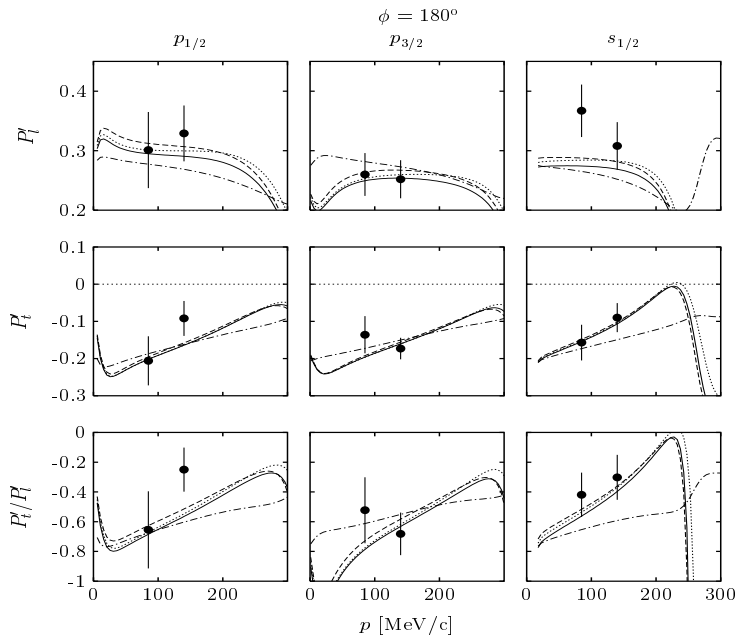,width=9.0cm,height=8.0cm}}
\caption{Transferred polarization asymmetries and their quotient
$P'_t/P'_l$ for proton knock-out from $^{16}$O for $q=1000$ MeV/c and
$\omega=450$ MeV. The electron energy is $\epsilon_e=2450$ MeV and 
 $\phi=180^{\rm o}$. Dotted lines: DWIA
with OB current only; Solid: total OB+MEC result;
Dot-dashed: The OB+MEC result but without the spin-orbit term of the
optical potential. All of them have been obtained using the
electromagnetic nucleon form factors of Galster.  Dashed: The total
OB+MEC result using instead the form factors of Gari-Krumplemann.}
\label{malov}
\end{figure}

Coming back to the case of polarization observables,
in  Fig.\ref{malov} we compare the computed transferred polarization with
the recent experimental data from TJlab\cite{Mal00}.
Although being aware of the possible modifications that the
``dynamical'' relativistic ingredients may introduce in the present
calculations,  the results in this figure
give us a clear indication of how much the DWIA calculation is
expected to be modified after including the two-body (MEC)
contributions (compare dotted with solid lines).  We see that the
effects of MEC lead to a global reduction of all of these polarization
observables, hence the OB calculation using the 
Gari-Krumplemann form factors would
be clearly located above the corresponding results including MEC
(dashed lines). This makes our present results to come closer to the
relativistic ones\cite{Mal00}.  Note also that the uncertainty
introduced by the nucleon form factor parameterization shows up in
$P'_l$, being negligible for $P'_t$.  

\begin{figure}[htb]
\centerline{
\epsfig{file=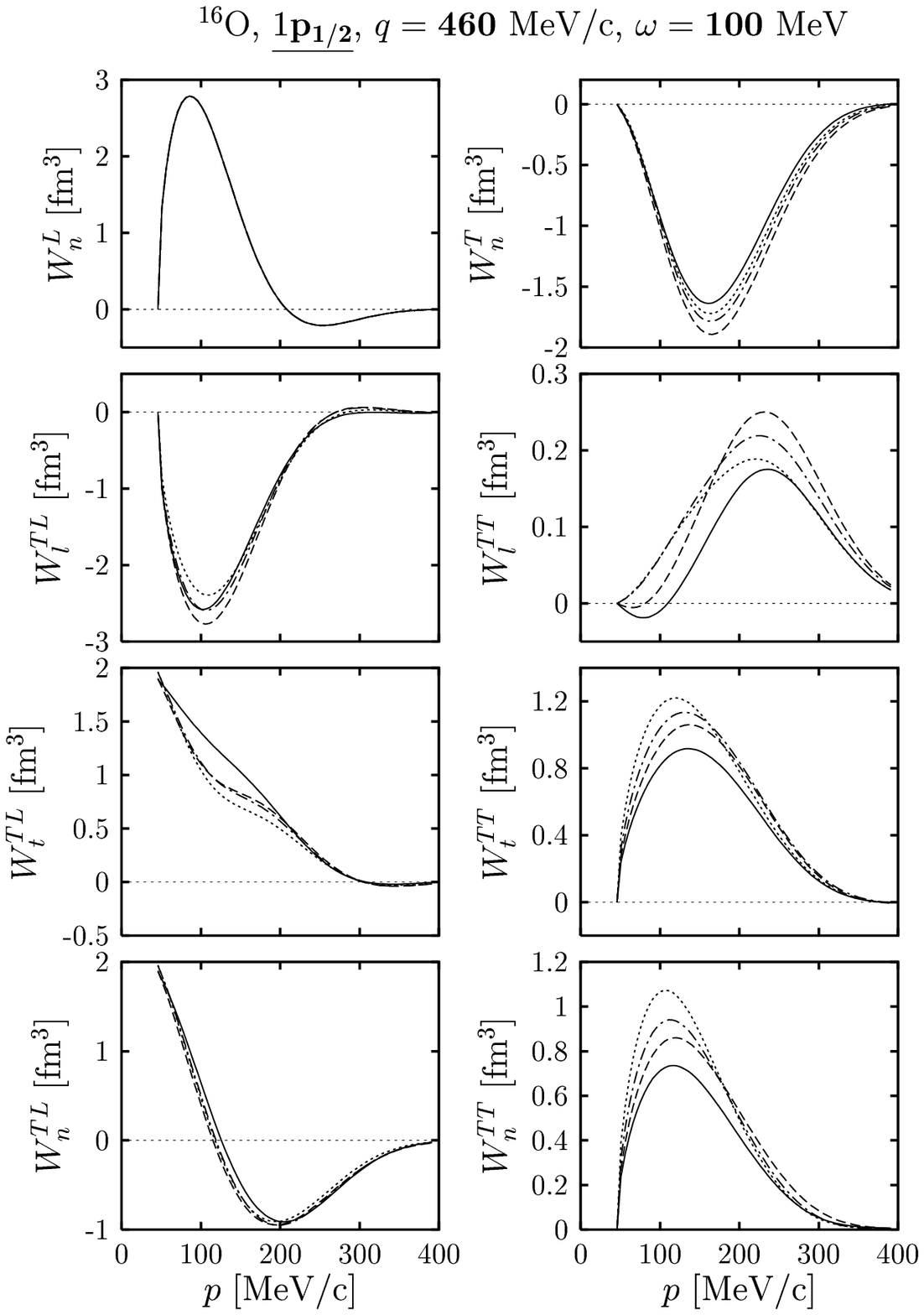,width=6cm,height=8.0cm}
\epsfig{file=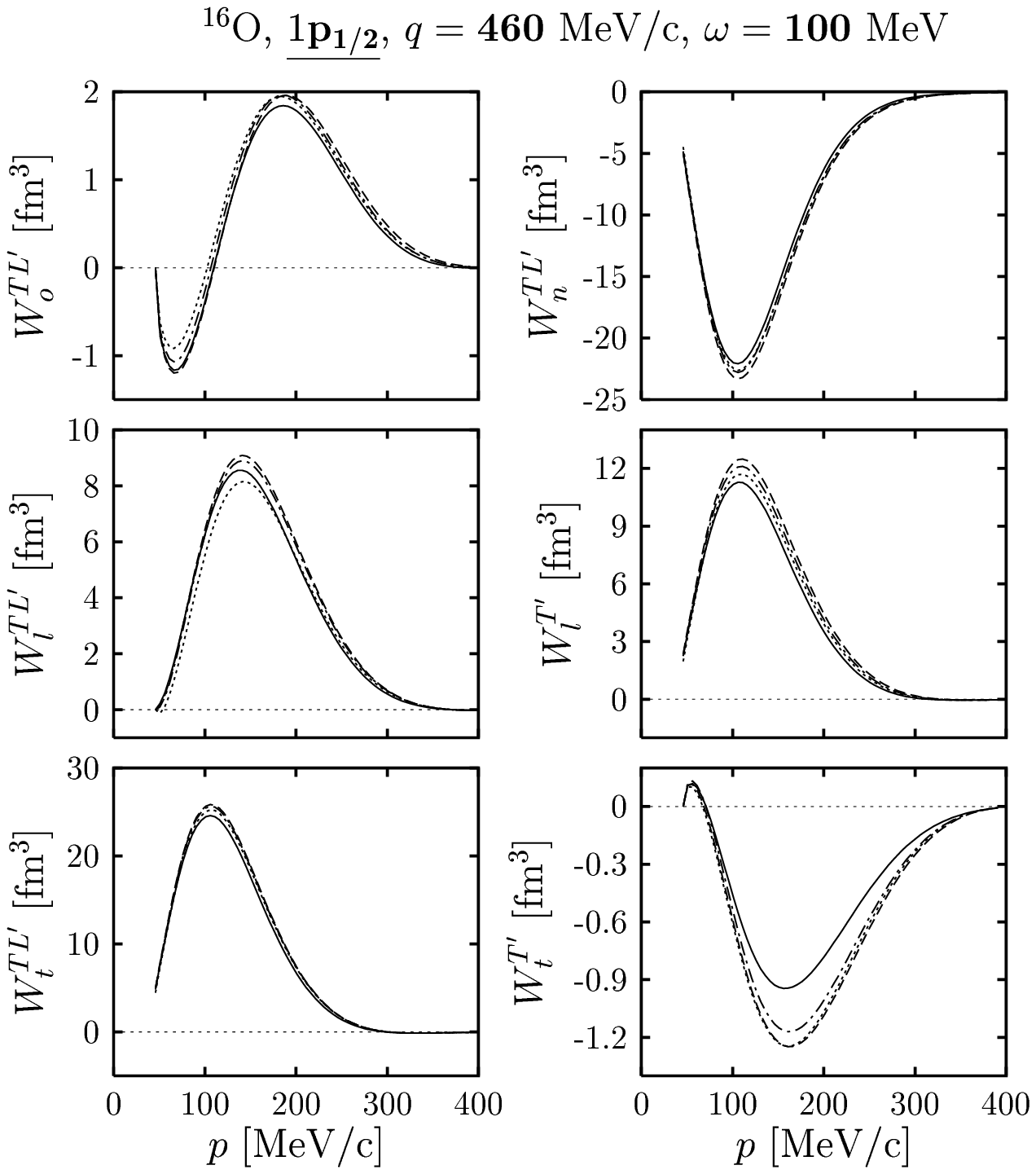,width=6cm,height=8.0cm}}
\caption{
The polarized response functions and fifth response
function for $p_{1/2}$ proton knock-out in $^{16}$O.
Dotted: DWIA using only the OB current operator; Dashed:
OB+Seagull; Dot-dashed: OB+Seagull+Pionic; Solid: total result
(OB+MEC) including also the $\Delta$ current.}
\label{W}
\end{figure}

Finally in Fig.\ref{W} we show an example of what typically we find
regarding the effects of MEC on the 13 polarized response functions and
the ``fifth'' response function\cite{Kaz03} for the $1p_{1/2}$ shell
of $^{16}$O for intermediate momentum transfer.  In general, MEC
effects over the transferred $T'$, $TL'$ responses are small and tend
to increase as $q$ goes higher\cite{Kaz03}, due mainly to the $\Delta$
current.  The role of MEC gets clearly more important for the induced
$T$, $TL$ and $TT$ polarized responses. Emphasis should be placed on
$W^{TT}_t$ and $W^{TT}_n$ (Fig.\ref{W}) which are reduced at the
maximum by $\sim 20\%$ and $\sim 30\%$, respectively. However these
effects are negligible in the case of $p_{3/2}$. For $q=1$ GeV/c the
role of MEC diminishes\cite{Kaz03}.

In conclusion, our results are showing that MEC effects are in general
small over the $(e,e'p)$ observables for low missing momentum,
although they are appreciable over some of the separate polarized
response functions.  Non negligible MEC effects are found for higher
missing momentum ($p>300$ MeV/c) over the asymmetries and
polarizations.  Note however that in this region dynamical
relativistic effects are also essential for describing the process, so
it would be desirable to have a completely relativistic model
including MEC in order to describe properly the high missing momentum
dependence.


\end{document}